\documentclass[pra, aps]{revtex4}
\usepackage{amsmath,amssymb}
\usepackage{psfrag}
\usepackage[dvips]{epsfig}
\usepackage{epsfig}

\usepackage{theorem}

\theorembodyfont{\rmfamily}

\theoremstyle{break}
\newtheorem{example}{Example}
\def\QED{~\rule[-1pt]{5pt}{5pt}\par\medskip}

\DeclareMathOperator{\tr}{tr}

\def\C{\mathbb{C}}
\def\Z{\mathbb{Z}}

\def\SU{\text{SU}}  
\def\U{\text{U}}  

\begin{document} 

\title{Generation of quantum logic operations from physical Hamiltonians}

\author{Jun Zhang$^{1, 2}$ and K. Birgitta Whaley$^{1}$}

\affiliation{$^1$Department of Chemistry and Pitzer Center for Theoretical Chemistry, 
University of California, Berkeley, CA 94720\\
$^2$Department of Electrical Engineering  
and Computer Sciences, University of California, Berkeley, CA 94720} 

\date{\today}  

\begin{abstract}
  We provide a systematic analysis of the physical generation of
  single- and two-qubit quantum operations from Hamiltonians available
  in various quantum systems for scalable quantum information
  processing.  We show that generation of one-qubit operations can be
  transformed into a steering problem on the Bloch sphere, whereas the
  two-qubit problem can be generally transformed into a steering
  problem in a tetrahedron representing all the local equivalence
  classes of two-qubit operations (the Weyl chamber). We use this
  approach to investigate several physical examples for the generation
  of two-qubit operations. The steering approach provides useful
  guidance for the realization of various quantum computation schemes.
\end{abstract}

\maketitle 

\section{Introduction}
Physical implementation of quantum information processing and quantum
computation usually begins with a coupled quantum mechanical system
and requires that the available Hamiltonian for that system be
controlled to generate desired quantum operations from which the
quantum algorithm is constructed~\cite{Barenco:95, Nielsen:00}. It is
therefore a fundamental issue to produce quantum operations from
the Hamiltonian that is provided by or accessible to the physical
quantum system.  In this work we formulate the generation of these
operations as a problem in control theory and show that this language
is beneficial to developing efficient physical implementation of
quantum operations for computation and information processing.

The postulates of quantum mechanics assert that the state of a quantum
system is completely described, at time $t$, by a unit vector
$|\psi(t)\rangle$ in a Hilbert space~\cite{Sakurai}. The evolution of
$|\psi(t) \rangle$ is determined by $|\psi(t)\rangle=
U(t)|\psi(0)\rangle$, where $U(t)$ is a unitary evolution operator (or
propagator). The dynamics of $U(t)$ is given by Schr\"{o}dinger's
equation $i\hbar\dot U(t)=H(v)U(t)$ with $U(0)=I$, where $H$ is the
Hamiltonian of the system, and $v$ the external control field. The
generation of quantum operation means to find a control $v$ such that
the trajectory generated by Schr\"{o}dinger's equation can achieve a
prescribed target unitary operator $U_T$ at certain final time.  This
can be posed as a steering problem in control theory.  The state space
of this control system is the unitary Lie group $\U(2^n)$, which
consists of all the quantum operations on $n$-qubit system, and the
dynamics of quantum operation described by Schr\"{o}dinger's equation
is a right invariant vector field on the unitary Lie group $\U(2^n)$.

Steering on Lie groups has been studied for a long time, and it still
remains a very difficult problem in control theory~\cite{sastry:99}.
It is well-known that when the control distribution is bracket
generating, the system is controllable, i.e., there exists a control
law that steers the system from any initial state to any final state
in the state space~\cite{Jur72, Jurdevic:97}. Nevertheless, this result
does not provide a constructive way to find the control law.  One
useful mathematical tool to solve steering problem is to use
Pontryagin's maximum principle, which gives a set of differential
equations that have to be satisfied by the optimal
trajectory~\cite{Bai78, DAl01, DAl02, Khaneja:01}.  Quite often however,
this approach leads to a two-point fixed-boundary problem which
has no analytic solutions~\cite{Bryson:75}.

In this paper, we develop a general approach to design the control law
for generation of an arbitrary quantum operation from the available
Hamiltonians in quantum mechanical systems. Because any quantum
operation on any arbitrary large quantum system can be decomposed as a
combination of single- and two-qubit quantum
operations~\cite{Barenco:95}, we only need to study the control of
single two-level systems and pairs of coupled two-level systems, which
amounts to the steering problem on the unitary Lie groups $\U(2)$ and
$\U(4)$ respectively. The key idea of the approach here is based on
the observation that in many real quantum systems there exists a
certain subset of quantum operations that can be readily generated.
This observation leads naturally to the notion of control on a quotient
space, that is, instead of controlling the original high dimensional
complicated quantum system, we study a reduced control problem on a
quotient space obtained from an equivalence relation that is defined
for two quantum operations if they differ only by quantum operations
in that specific subset.  Therefore, to generate a desired target
quantum operation, we can equivalently generate a quantum operation
that differs from the target by a quantum operation in the easily
achieved subset. One key issue in this approach is to find the
structure of the quotient space and the reduced control trajectory on
this quotient space. This makes it possible to study the generation of
quantum operations by solving a steering problem on the quotient
space.  We present solutions here for steering on the quotient space
of both $\U(2)$ and of $\U(4)$.  
The solutions for $\U(2)$ are related to those known from the NMR
literature, but are developed here in a unified steering framework that also
encompasses the 
solutions for the more complex $\U(2)$.  

We start with the generation of single-qubit operations, which can be
viewed as a control problem on a two-level system. We know that all
the single-qubit operations form the unitary Lie group $\U(2)$, and that
any single-qubit operation can be described as a rotation on the Bloch
sphere of a certain angle about a specified axis. In many two-level
quantum systems, it is often easy to generate a rotation operator
$R_{\hat n}$ about a fixed axis $\hat n$ by appropriate choice of a static 
control
field.  By making a change of coordinates one can transform $R_{\hat n}$
into $R_z$, i.e., the rotation operator about the $z$-axis.  Consequently, to
implement an arbitrary single-qubit quantum operation $U_T$, we only
need to implement a quantum operation $U_1$ that differs from $U_T$ by
a $z$ rotation operator from the right, that is, $U_1=U_T
R_z(\theta)$, with $\theta\in [0, 2\pi)$. This defines an equivalence
relation on all the single-qubit quantum operations, and the quotient
space obtained from this equivalence relation is nothing but the Bloch
sphere. Therefore, the generation of single-qubit operations can be
reduced to a steering problem on the Bloch sphere.  For the case when
the control is an oscillating field perpendicular to the static field,
we derive the control laws for generation of arbitrary single-qubit
operations that are familiar from the NMR literature.  We also prove
that the system is controllable if and only if the oscillating field
is tuned at the resonant frequency. We then study the case when the
oscillating field is not perpendicular to the static field. For both
of these two cases, we show that when at resonant frequency, the total
time to implement a given operation is quantized. This is a
particularly important issue for the implementation of local unitary
operations in coupled two-qubit systems, for which two
single-qubit operations must be generated at the same time. We also show
that when the control variable is varied within a certain range, an
efficient way to generate any single-qubit operation is to simply
alternate between the two extremal values of the control parameter.
A control law design based on the Bloch sphere steering technique
thus provides a complete solution to the generation of any arbitrary
local unitary operations from a given Hamiltonian.

We then consider the generation of two-qubit quantum operations.
Physically, this is a control problem on two coupled two-level
systems. It is much more difficult to control a coupled qubit system
to generate the desired two-qubit quantum operation, because of the
higher dimensionality and more complicated Lie group structure.  In
Ref.~\cite{Khaneja:01}, Khaneja {\it et al.} derived the time optimal
solution when the Hamiltonian is purely nonlocal. However, when the
Hamiltonian contains both local and nonlocal terms with comparable
magnitudes, there are no systematic methods to derive the control
laws other than numerical procedures~\cite{Niskanen:03, Schulte:04}.
In this paper, we provide a systematic approach to solve this problem
by using the notion of control on a quotient space. Our control strategy
employs two steps to generate an arbitrary two-qubit operation $U_T$.
First, we find an efficient control law to generate a nonlocal quantum
operation $U_1$ that differs from $U_T$ only by local unitary
operations $k_1$ and $k_2$, where each of these is the tensor
product of two single-qubit quantum operations. Then we consider
generation of the local operations $k_1$ and $k_2$ that is required
in order to arrive at $U_T$ in the computational basis.  This will often
require that the single-qubit operations be implemented simultaneously 
(see above).

A basic principle in this control design strategy is that we implement the
local and two-qubit operations separately. This is often feasible in
many physical applications, where the single-qubit operations are
generated by local Hamiltonians that can be independently controlled,
e.g., by external fields, so that we can readily implement any local
unitary operation.  This separation leads naturally to the concept of local
equivalence on all two-qubit quantum operations~\cite{Makhlin:00}.
In Ref.~\cite{Zhang:02}, we derived the geometric representation of
this equivalence relation as a tetrahedron, referred to as the Weyl
chamber.  Each point in the tetrahedron represents a local equivalence
class of some nonlocal two-qubit operations.  The generation of a
two-qubit quantum operation can then be treated as a steering problem
in this tetrahedron. We have shown some examples of this steering
technique~\cite{Zhang:02}, when the Hamiltonian is given in some
special cases, e.g., purely nonlocal.  In this paper, we will exploit
this idea to more systematically study the generation of two-qubit
operations for a variety of Hamiltonians that include both nonlocal and local
terms and that derive from physical
systems relevant to implementation of quantum information processing, in 
particular to solid state implementations.

\section{Generation of single-qubit operations: steering on Bloch sphere}
\label{sec:single}
In this section, we study the generation of single-qubit operations 
as a control problem of unitary transformations on two-level quantum systems. 
The time evolution 
operator for a
general two-level system can be written as follows:
\begin{equation}
\label{eq:t1}
  i\dot U = (H_d + v H_c) U, \quad U(0)=I,
\end{equation}
where $H_d$ is the drift Hamiltonian, $H_c$ the control Hamiltonian,
and $v$ the control. Here $H_d$ and $H_c$ are both Hermitian
matrices on $\C^{2\times 2}$.  Upon neglecting the global phase term, all
single-qubit quantum operations form a Lie group $\SU(2)$:
\begin{equation*}
  \SU(2)=\{ U\in \C^{2\times 2} : UU^\dag = I, \det U=1\}.
\end{equation*}
Our task is to find the control $v$ that will drive the
system~\eqref{eq:t1} from the initial operation $U(0)=I$ to a
prescribed target quantum operation $U_T$.  Note that we have implicitly
subdivided the physical Hamiltonian into a term that cannot be varied,
$H_d$, and a term that may be controlled externally, $H_c$.

To solve this control problem, we first transform Eq.~\eqref{eq:t1} into the
following standard form:
\begin{equation}
  \label{eq:t2}
  i\dot U = (a\sigma_z + v H_c) U.
\end{equation}
This transformation is obtained by the following argument.
Without loss of generality, we can assume that $H_d
=a_1\sigma_x+a_2\sigma_y+a_3\sigma_z$, where $\sigma_x$, $\sigma_y$,
and $\sigma_z$ are Pauli matrices. When $a_1^2+a_2^2\neq 0$, let
\begin{equation*}
  k=\left(
    \begin{matrix}
    \dfrac{a_1-a_2 i}{\sqrt{2a(a-a_3)}} & \dfrac{-a_1+a_2
    i}{\sqrt{2a(a+a_3)}}\\
\sqrt{\dfrac{a-a_3}{2a}}&\sqrt{\dfrac{a+a_3}{2a}}
    \end{matrix}
\right),
\end{equation*}
where $a=\sqrt{a_1^2+a_2^2+a_3^2}$.  It is straightforward to verify
that $kk^\dag=I$, i.e., $k$ is a single-qubit operation. One can also verify
that $kH_dk^{-1}=a\sigma_z$.  Then, letting $U_1=kU$, we have
\begin{eqnarray*}
\aligned
i\dot U_1&= k(H_d+vH_c)k^{-1} U_1\\
&=(a\sigma_z+vkH_ck^{-1})U_1.  
\endaligned
\end{eqnarray*}
where we have recognized that $kH_ck^{-1} = H'_c$, which puts this
into the form of Eq.~(\ref{eq:t2}).  The original control problem of
steering Eq.~\eqref{eq:t1} from $U(0)=I$ to $U_T$ is thereby
transformed into a problem of steering Eq.~\eqref{eq:t2} from
$U_1(0)=k$ to $kU_T$. For control of a single qubit, we therefore only
need to consider the standard form~\eqref{eq:t2}.

\subsection{Steering on Bloch sphere}
If no control signals are applied, i.e., $v=0$, Eq.~\eqref{eq:t2} simplifies to
\begin{equation}
\label{eq:t3}
  i\dot U = a \sigma_z U.
\end{equation}
The solution to Eq.~\eqref{eq:t3} is simply the $\sigma_z$ rotation
operator $R_z(2at)=e^{-ia\sigma_z t}$. We can thus easily implement any
$\sigma_z$ rotation by turning off the control $v$ and letting the
system evolve for a certain time duration under $H_d$ alone. 
We now define the notion of $R_z$-equivalence, namely, that
$U_1$ and $U_2$ are $R_z$-equivalent if they satisfy
\begin{equation}
  \label{eq:t4}
  U_T=U_1 e^{-i\sigma_z t_1}.
\end{equation}
It is evident that this relation is reflexive, symmetric, and
transitive, and therefore it is an equivalence relation on the Lie
group $\SU(2)$ of all the single-qubit operations. 
To generate an arbitrary single-qubit operation $U_T$, it is then
sufficient to generate an operation $U_1$ that differs from $U_T$ by a
$\sigma_z$ rotation from the right.  Now our task has been reduced to
a control problem on the space of all the $R_z$-equivalence classes,
that is, on a quotient space of $\SU(2)$.

\begin{figure}[tb]
\begin{center}
 \psfrag{x}[][]{$x$}
 \psfrag{y}[][]{$y$}
 \psfrag{2}[][]{$\theta$}
 \psfrag{3}[][]{$\phi$}
  \psfrag{0}[][]{$|0\rangle$}
 \psfrag{1}[][]{$|1\rangle$}
 \psfrag{4}[][]{$|\psi\rangle$}
\includegraphics[width=0.25\hsize]{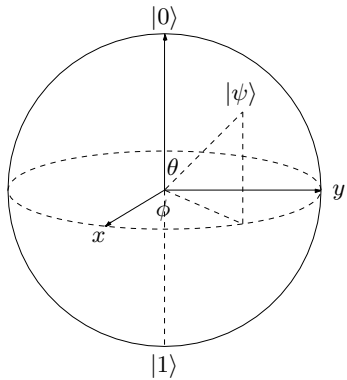}
\end{center}
\caption{Bloch sphere representation of a single-qubit.}
\label{fig:bloch}
\end{figure}

We now show that this quotient space is nothing but the Bloch sphere.
Recall that the state of a single qubit can be represented by $|\psi
\rangle=z_1|0\rangle+z_2|1\rangle$, where $z_i$ are complex numbers
with restriction $|z_1|^2+|z_2|^2=1$.  Ignoring the global phase,
this state can be written as
 \begin{equation*}
    |\psi\rangle=\cos\frac{\theta}2|0\rangle
 +e^{i\phi}\sin\frac\theta{2}|1\rangle,
 \end{equation*}
 where the real numbers $\theta$ and $\phi$ define a point $(x, y,
 z)=(\sin\theta\cos\phi, \sin\theta\sin\phi, \cos\theta)$ on a unit
 sphere, which is the well-known \emph{Bloch sphere} shown in
 Fig.~\ref{fig:bloch}.  To find $\theta$ and $\phi$ explicitly, we can
 use the Hopf fibration, which is a map $\pi: S^3\to S^2$ defined by
 $\pi(\psi)=(x, y, z) \in S^2$~\cite{Marsden:99}, with
 \begin{eqnarray}
\label{eq:hopf}
   (z, x+yi)=(|z_1|^2 - |z_2|^2, 2{\bar z_1} z_2).
\end{eqnarray}
We define a map $\Phi$ from a single-qubit operation to a point on the
Bloch sphere by $\Phi(U)=U|1\rangle$, where $|1\rangle$ is the
single-qubit state corresponding to the south pole on the Bloch
sphere.  Then
\begin{equation*}
  \Phi(e^{-ia\sigma_z t})=e^{-ia\sigma_z t} |1\rangle=|1\rangle.
\end{equation*}
Furthermore, if
$U_1$ and $U_2$ are $R_z$-equivalent, we have $\Phi(U_1)=\Phi(U_2)$.
Therefore, all operations that are $R_z$-equivalent are mapped to
the same point on the Bloch sphere.  Conversely, it can also be proved
by Lie group theory~\cite{Marsden:99} that all operations that are
mapped to the same point on the Bloch sphere differ only by a
$\sigma_z$ rotation from the right, that is, they are all
$R_z$-equivalent. We thereby obtain that the quotient space of
$R_z$-equivalence classes on $\SU(2)$ can be represented by the Bloch
sphere.  

Now our control task becomes a steering problem on the Bloch sphere.
Under the above mapping, the initial point of this steering problem is
$\Phi(I)=|1\rangle$, and the final point is $U_T|1\rangle$, where
$U_T$ is the desired target operation. Our strategy is to first find
the control $v$ that can generate a trajectory on the Bloch sphere
reaching the final point $U_T|1\rangle$ from the initial point
$|1\rangle$. Applying this control to the single-qubit
system~\eqref{eq:t2}, we can obtain an operation $\tilde U_T$ that is
$R_z$-equivalent to the target operation $U_T$. The second step is to
find the appropriate additional time duration $t_z$ such that
$U_T=\tilde U_T e^{-ia\sigma_z t_z}$.  The detailed solutions will depend on 
the form of the control term $v H_c$.  

\subsection{Oscillating control field}
We illustrate this approach of control on the quotient space with
analysis of generation of single-qubit operations on two-level quantum
systems subject to an oscillating electromagnetic control field $A/2
\cos(\omega t+\delta)$.  Some of the results shown below are known
within the NMR literature, in particular that for a control field
perpendicular to the static field. However, the solutions look
different because the analysis here is made in the laboratory frame,
instead of in a rotating frame as is customary in the NMR community.
In NMR, under usual operating conditions the coupling terms simplify
to an Ising interaction which is invariant under the rotating frame
transformation. In the general case, when the coupling term is not of
the Ising form, a time independent coupling Hamiltonian in the
laboratory frame will be time dependent in the rotating frame. As
noted above, implementation of a two-qubit local gate, i.e.,
$k_1=k_{11}\otimes k_{12}$, requires simultaneous implementation of
each single-qubit operation $k_{11}$ and $k_{12}$.
We show below that such simultaneous single-qubit operations are
readily steered in the laboratory frame. The same argument applies to
local gates in larger multi-qubit arrays.  

\subsubsection{Perpendicular control field}
We first consider the case in which the control Hamiltonian $H_c$ is
an oscillating control field perpendicular to a static drift field
$H_d$.  This is prevalent in many quantum systems, e.g., Electron Spin
Resonance (ESR).  Here a single atom is placed in a uniform constant
magnetic field directed along the $z$-axis, and the control is an
oscillating magnetic field oriented along the
$x$-axis~\cite{Bransden}. The dynamics of the quantum spin is
determined by:
\begin{equation}
  \label{eq:t5}
i\dot U=\bigg(\frac{\omega_0}2 \sigma_z+\frac{A }2 \cos(\omega
  t+\delta) \sigma_x\bigg) U.
\end{equation}

A solution to Eq.~\eqref{eq:t5} for near resonance, $\omega \sim
\omega_0$, may be found by transforming first to a rotating frame
defined by the drift (Larmor) frequency, i.e., $U_1=e^{i\omega_0 t/2
  \sigma_z}U$. We then obtain
\begin{eqnarray*}
  i\dot U_1&=&\frac{A}2 \cos (\omega t+\delta)
\big(\sigma_x \cos \omega_0 t -\sigma_y \sin \omega_0 t\big) U_1\\
&=& \frac{A}4 \left(
  \begin{matrix}
  & e^{i[(\omega+\omega_0)t+\delta]} +e^{-i[(\omega-\omega_0)t+\delta]}\\
 e^{-i[(\omega+\omega_0)t+\delta]} +e^{i[(\omega-\omega_0)t+\delta]}&\\ 
  \end{matrix}
\right) U_1.
\end{eqnarray*}
Since for near resonance between drift and control fields the terms
$e^{\pm i[(\omega+\omega_0)t+\delta]}$ oscillate much faster than the
terms $e^{\pm i[(\omega-\omega_0)t+\delta]}$, they make on average
little contribution to $\dot U_1$ and can be neglected (rotating wave
approximation), resulting in the approximate equation:
\begin{equation}
  \label{eq:t6}
  i\dot U_1= \frac{A}4\bigg(
\cos[(\omega-\omega_0)t+\delta]\sigma_x+\sin[(\omega-\omega_0)t+\delta]\sigma_y
 \bigg)U_1.
\end{equation}
Now letting $U_2=e^{i[(\omega-\omega_0)t+\delta]/2 \sigma_z}U_1$, we have
\begin{equation}
 \label{eq:t7}
  i\dot U_2=\bigg(\frac{A}4 \sigma_x+\frac{\omega_0-\omega}2\sigma_z\bigg)U_2,
\end{equation}
which has an explicit solution
\begin{equation*}
  U_2(t)=e^{-i({A}/4 \sigma_x+(\omega_0-\omega)/2\sigma_z)t}.
\end{equation*}
Combining these two transformations, we obtain an approximate solution 
(i.e., within the rotating wave approximation) to Eq.~\eqref{eq:t5} as
\begin{equation}
  \label{eq:t8} 
U(t_f)=e^{-i(\omega t_f+\delta)/2 \sigma_z}
e^{-i({A}/4 \sigma_x+(\omega_0-\omega)/2\sigma_z)t_f}
 e^{i{\delta}/2 \sigma_z}.
\end{equation}
This is equivalent to the standard treatment of ESR and NMR~\cite{Bransden}.
Since we can also generate any $\sigma_z$ rotation by turning off the
control field $H_c$ for a time duration $t_z$, a general form of the quantum
operation generated by this system can then be written as
\begin{equation}
  \label{eq:t8a}
 U(t_f, t_z)=U(t_f) e^{-i\omega_0/2\sigma_z t_z} =
 e^{-i(\omega t_f+\delta)/2 \sigma_z}
e^{-i({A}/4 \sigma_x+(\omega_0-\omega)/2\sigma_z)t_f}
 e^{i(\delta- \omega_0 t_z )/2\sigma_z} . 
\end{equation}

\begin{figure}[tb]
\begin{center}
\begin{tabular}{cc}
 \psfrag{x}[][]{$x$}
 \psfrag{z}[][]{$z$}
\psfrag{1}[][]{$|1\rangle$}
 \psfrag{n}[][]{\footnotesize$(\frac{A}2, \omega_0-\omega)$}
 \psfrag{m}[][]{$z=\frac{(A/2)^2-(\omega_0-\omega)^2}
{(A/2)^2+(\omega_0-\omega)^2}$}
\includegraphics[width=0.32\hsize]{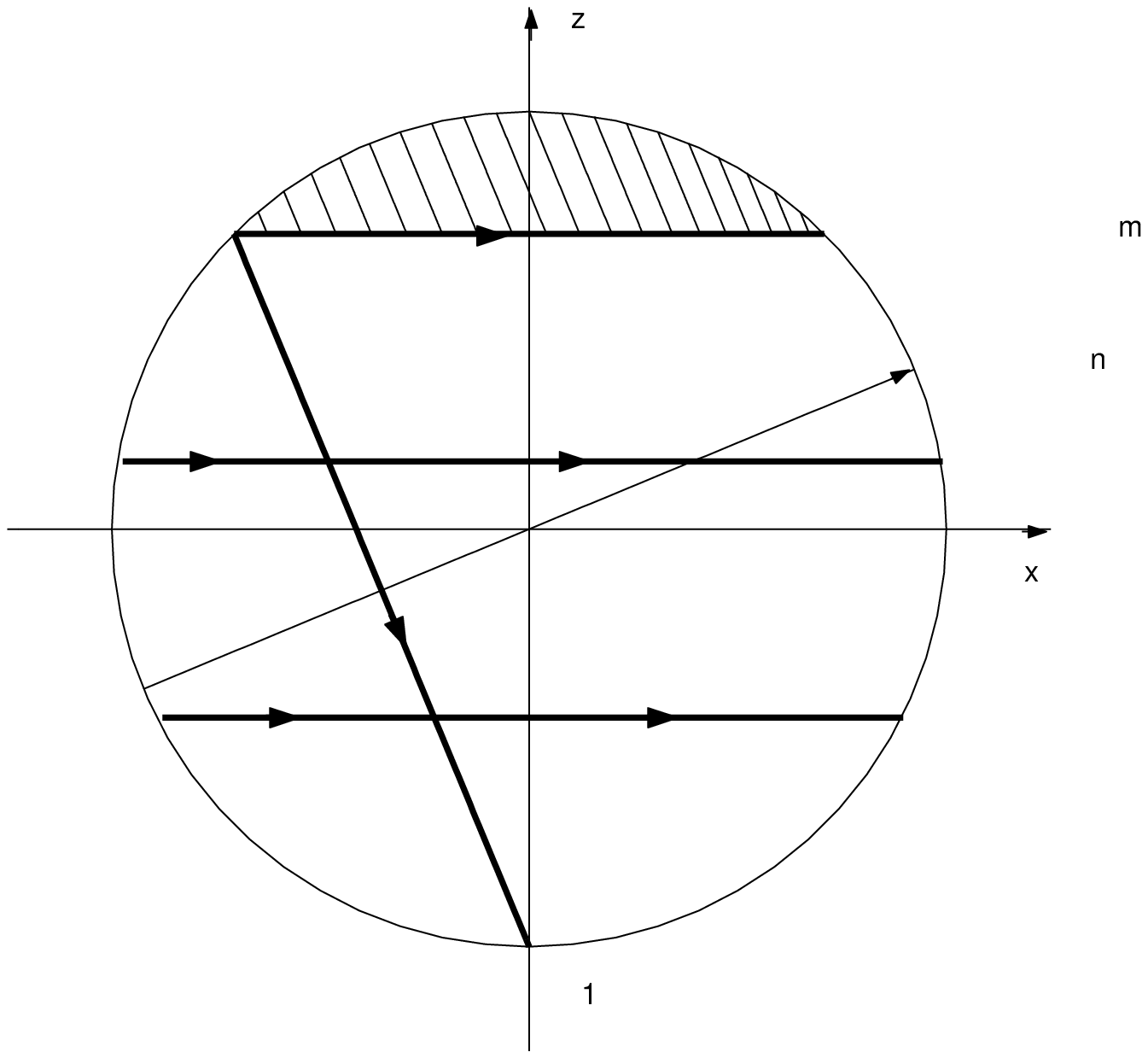}
&\qquad \qquad
 \psfrag{x}[][]{$x$}
 \psfrag{z}[][]{$z$}
\psfrag{1}[][]{$|1\rangle$}
\includegraphics[width=0.3\hsize]{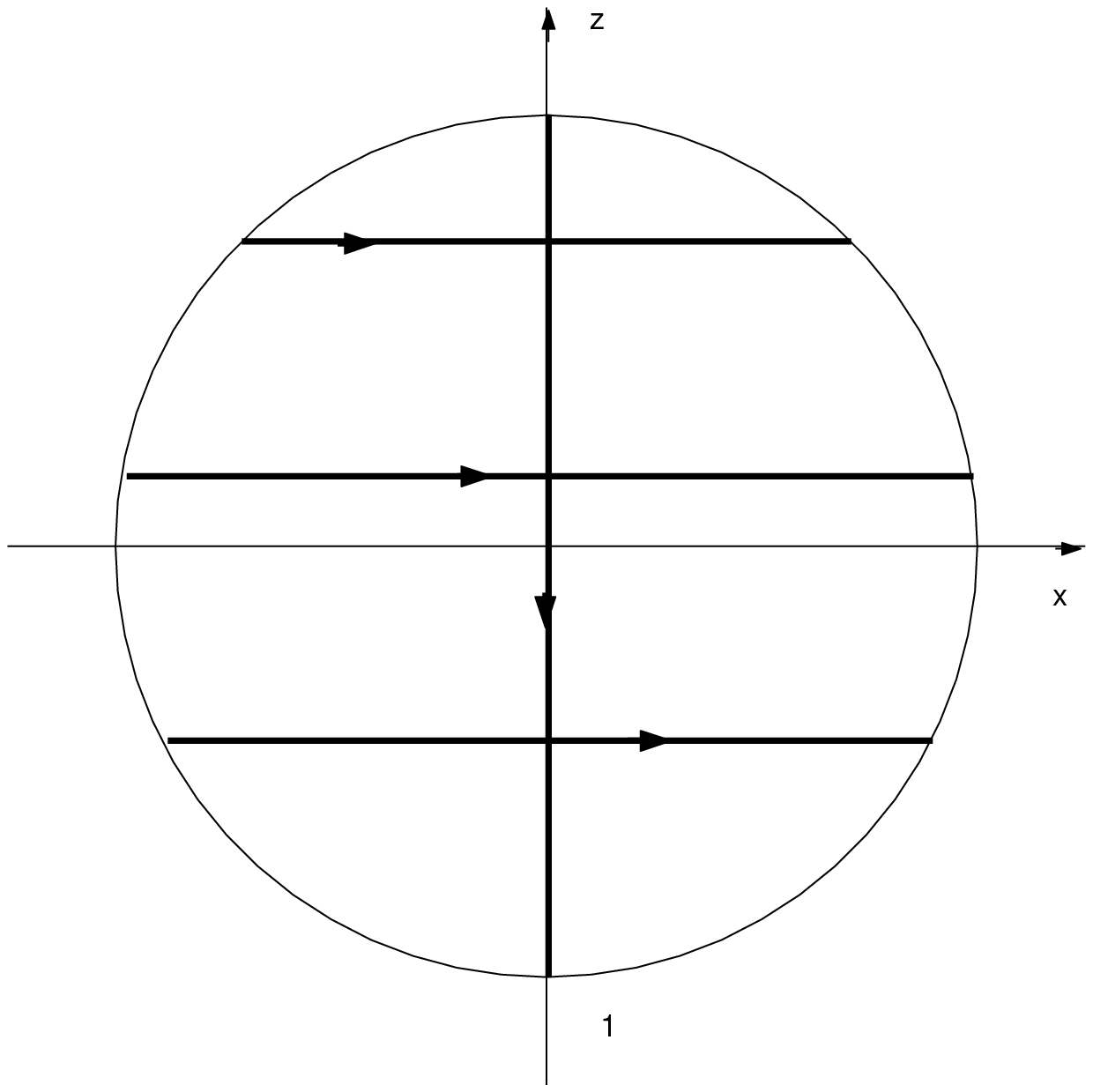}\\
(A)&\qquad \qquad(B)
\end{tabular}
\end{center}
\caption{The effect of the single-qubit operation
$U(t_f, t_z)$, Eq.~(\ref{eq:t8a}), on the quantum state $|1\rangle$:
(A) $\omega\neq\omega_0$; (B) $\omega=\omega_0$.}
\label{fig:Ut}
\end{figure}

In the language of control on the quotient space developed above, the
effect of the operation $U(t_f, t_z)$ on the quantum state $|1\rangle$
can be described directly from Eq.~\eqref{eq:t8a} as a rotation about
the axis $(A/2, 0, \omega_0-\omega)$ followed by another rotation
about the $z$-axis, as shown in Fig.~\ref{fig:Ut} for both near resonant (A)
and resonant (B) situations. From this it is evident
that when $\omega\neq \omega_0$, all points $(x, y, z)$ on the Bloch
sphere satisfying
\begin{equation}
  \label{eq:unctrl}
z\ge   \frac{(A/2)^2-(\omega_0-\omega)^2}{(A/2)^2+(\omega_0-\omega)^2},  
\end{equation}
shown as the shaded area in Fig.~\ref{fig:Ut},
can never be reached for any choice of
tuning parameters $A$, $\delta$ and time durations $t_f$, $t_z$.
Consequently, the reachable set in this near-resonant case is not {\it all}
single-qubit operations. To allow implementation of any arbitrary single-qubit
operation, we must impose the strict condition $\omega=\omega_0$, that
is, the oscillating field must be tuned to the resonant frequency.

We now show that when operating at exact resonance, the time
duration to implement an arbitrary single-qubit rotation in the laboratory
frame is quantized.
At resonance, Eq.~\eqref{eq:t8a} becomes
\begin{equation}
  \label{eq:t9} 
U(t_f, t_z)=e^{-i(\omega_0 t_f+\delta)/2 \sigma_z} e^{-i {A}/4 \sigma_x t_f}
 e^{i(\delta- \omega_0 t_z )/2\sigma_z}. 
\end{equation}
On the other hand, we know from Euler's ZXZ decomposition that an
arbitrary single-qubit operation $U_T$ can be decomposed as
\begin{equation}
  \label{eq:t10}
  U_T=e^{-i(\phi-\pi/2)/2\sigma_z} e^{-i(\pi-\theta)/2\sigma_x} 
e^{-i\gamma/2 \sigma_z}.
\end{equation}
Note that in Eq.~\eqref{eq:t10} the coefficients in the first two
terms are chosen such that $U_T$ is mapped to the point $(\theta,
\phi)$ on the Bloch sphere.  Comparing Eqs.~\eqref{eq:t9}
and~\eqref{eq:t10}, we find that for $\omega = \omega_0$, an arbitrary
single-qubit operation $U_T$ in Eq.~\eqref{eq:t10} can be implemented
if the following conditions are satisfied:
\begin{eqnarray}
 \label{eq:t11a}
{\phi-\pi/2}+2m_1\pi&=& {\omega_0 t_f+\delta},\\
\label{eq:t11b}
{\pi-\theta}+2m_2\pi&=&\frac{A}2 t_f,\\
\label{eq:t11c}
-\gamma+2m_3\pi&=&{\delta}-{\omega_0} t_z,
\end{eqnarray}
where $m_1$, $m_2$, $m_3\in \Z$, and $m_1+m_2+m_3$ is even.  Any
values of the control parameters $A$ and $\delta$, and time durations
$t_z$ and $t_f$ that satisfy Eqs.~\eqref{eq:t11a}--\eqref{eq:t11c} can
generate the desired target operation $U_T$ within the total time
$t_f+t_z$. From Eqs.~\eqref{eq:t11a} and ~\eqref{eq:t11c}, we also
find that
\begin{equation}
 \label{eq:t12}
  t_f+t_z=\frac{\phi+\gamma-\pi/2+2m\pi}{\omega_0},
\end{equation}
where $m$ is an integer. Hence, from Eq.~\eqref{eq:t12}, it is evident
that if the oscillating field is tuned at resonant frequency, the time
duration to implement a single-qubit operation is quantized with a
time duration that is independent of the amplitude of the oscillating
control field. This is a particularly important issue in the
generation of two-qubit local unitary operations, which are given by
$k_1=k_{11}\otimes k_{12}$, where $k_{11}$ and $k_{12}$ are both
single-qubit operations. To implement such a two-qubit local unitary
$k_1$, we need to implement both $k_{11}$ and $k_{12}$ within the same
time.  We can satisfy this constraint by making the oscillating
control field that addresses one qubit to be resonant and that which
addresses the other qubit to be slightly off-resonance.  Using this
offset as an additional variable, we can then apply a numerical
optimization procedure to achieve $k_{11}$ and $k_{12}$
simultaneously, leading to the desired local two-qubit unitary.

The above analysis of the time duration is made in the laboratory
frame and thus differs from the standard NMR/ESR situation in which
pulses are made in the rotating frame~\cite{Lieven:04}
(Eq.~\eqref{eq:t12} applies only in the laboratory frame). When
working in the rotating frame $U_1=e^{i\omega_0 t/2 \sigma_z}U$, as is
usual in NMR experiments at resonance, the following time constraints
are instead required to implement $U_T$:
\begin{eqnarray}
 \label{eq:t13a}
 {\phi-\pi/2}+2m_1\pi&=&\delta,\\
 \label{eq:t13b}
{\pi-\theta}+2m_2\pi&=&\frac{A}2 t_f,\\
 \label{eq:t13c}
-\gamma+2m_3\pi&=&{\delta}-{\omega_0} t_z.
\end{eqnarray}
In this case, we have
\begin{equation*}
  t_f+t_z=\frac{2(\pi-\theta)+4m_2\pi}{A}+\frac{\phi+\gamma-\pi/2+2m\pi}
{\omega_0}.
\end{equation*}
Hence, when working in the rotating frame, it is possible to decrease the
time to reach a desired operation merely by increasing the amplitude of the oscillating field.

\subsubsection{Non-perpendicular control field}
We now consider the situation in which the control Hamiltonian is an
oscillating electromagnetic field that is not perpendicular to the
static field:
\begin{equation}
  \label{eq:t15}
 i\dot U_1=\bigg[\frac{\omega_0}2\sigma_z+\frac{{A}}2\cos(\omega
 t+\delta) (\cos \zeta \sigma_z -\sin\zeta \sigma_x)\bigg]U_1.
\end{equation}
Here the control Hamiltonian $H_c=\cos \zeta \sigma_z -\sin\zeta
\sigma_x$, so the static and oscillating fields are tilted by an
angle.  This can arise in NMR or in coupled semiconductor quantum
dots~\cite{Britton:04}.  Let $U_2=e^{i(\omega_0t+\delta)/2\sigma_z}
U_1$. Imposing the rotating wave approximation again, we have
\begin{equation}
  \label{eq:t16}
  i\dot U_2=\frac{{A}}4 \big[2\cos\zeta \cos(\omega
  t+\delta)\sigma_z -\sin\zeta(\cos(\omega-\omega_0)t\, \sigma_x
+\sin(\omega-\omega_0)t\, \sigma_y)\big]U_2.
\end{equation}
When the oscillating field $\omega$ is tuned at the resonant frequency $\omega_0$,
Eq.~\eqref{eq:t16} becomes
\begin{equation}
  \label{eq:t17}
  i\dot U_2=\frac{{A}}4 \big[2\cos\zeta \cos(\omega_0
  t+\delta)\sigma_z -\sin\zeta\sigma_x\big]U_2.
\end{equation}
Letting $U_3=e^{-i\pi/4\sigma_y}U_2$, we transform Eq.~\eqref{eq:t17}
into the standard form:
\begin{equation}
  \label{eq:t18}
  i\dot U_3=\frac{{A}}4 \big[2\cos\zeta \cos(\omega_0
  t+\delta)\sigma_x+\sin\zeta\sigma_z\big]U_3.
\end{equation}
Finally, letting $U_4=e^{i{A}\sin\zeta/4\sigma_z t}U_3$, we obtain:
\begin{eqnarray}
  \label{eq:t19}
\aligned
  i\dot U_4&=\frac{{A}}2 \cos\zeta \cos (\omega_0 t+\delta) \bigg[\sigma_x
  \cos\big(\frac{{A}}2 \sin\zeta t\big)  -\sigma_y 
\sin\big( \frac{{A}}2 \sin\zeta t\big) \bigg] U_4\\
&=\frac{{A}}4 \cos\zeta
\left(\begin{matrix}
  &e^{ i[(\omega_0-{A}/2 \sin\zeta)t+\delta]}
+e^{ i[(-\omega_0-{A}/2 \sin\zeta)t-\delta]}\\
e^{ i[(\omega_0+{A}/2 \sin\zeta)t+\delta]}
+e^{ i[(-\omega_0+{A}/2 \sin\zeta)t-\delta]}&
\end{matrix}\right) U_4.
\endaligned
\end{eqnarray}
Since the control fields are usually high frequency signals, we have
$\omega_0\gg {A}/2\sin\zeta$. Eq.~\eqref{eq:t19} can then be
simplified to
\begin{equation}
  \label{eq:t20}
  i\dot U_4=\frac{{A}}2 \cos\zeta \cos(\omega_0 t+\delta)\, \sigma_x U_4,
\end{equation}
which has a solution
\begin{equation}
  \label{eq:t21}
  U_4(t)=e^{-i{A}\cos\zeta[\sin(\omega_0t+\delta)-\sin\delta]
/(2\omega_0)\sigma_x}.
\end{equation}
Combining the above series of transformations, we obtain an approximate
solution to Eq.~\eqref{eq:t15} as:
\begin{equation}
  \label{eq:t22}
  U(t_f)=e^{-i(\omega_0 t_f +\delta)/2\sigma_z}
  e^{i{A}\sin\zeta/4\sigma_x t_f}
e^{-i [{A}\cos\zeta[\sin(\omega_0t_f+\delta)-\sin\delta]
/(2\omega_0)-\delta/2]  \sigma_z}.
\end{equation}
As in the previous example, we also have the freedom to implement any arbitrary $\sigma_z$
rotation by turning off the oscillating control field for a time
duration $t_z$.  Hence the final quantum operation achieved can be written
as:
\begin{equation}
  \label{eq:t23}
  U(t_f, t_z)=U(t_f) e^{-i\omega_0/2\sigma_z t_z} 
=e^{-i(\omega_0 t_f +\delta)/2\sigma_z}
  e^{i{A}\sin\zeta/4\sigma_x t_f}
e^{-i [{A}\cos\zeta[\sin(\omega_0t_f+\delta)-\sin\delta]
/(2\omega_0)-\delta/2 +\omega_0 t_z/2]  \sigma_z}.
\end{equation}

The time duration required to implement an arbitrary single-qubit operation 
under these conditions
may be similarly derived by comparing with the general form of $U_T$ given in
Eq.~\eqref{eq:t10}.  This results in the conditions:
\begin{eqnarray}
 \label{eq:t24a}
{\phi-\pi/2}+2m_1\pi&=& {\omega_0 t_f+\delta},\\
\label{eq:t24b}
{\pi-\theta}+2m_2\pi&=&-\frac{{A}}2 \sin\zeta t_f,\\
\label{eq:t24c}
-\gamma+2m_3\pi&=&-\frac{{A}\cos\zeta}{\omega_0}
[\sin(\omega_0t_f+\delta)-\sin\delta]
+\delta -\omega_0 t_z,
\end{eqnarray}
where $m_1$, $m_2$, $m_3\in \Z$, and $m_1+m_2+m_3$ is even. From
Eqs.~\eqref{eq:t24a} and~\eqref{eq:t24c}, we then find that
\begin{equation}
  \label{eq:t25}
t_f+t_z=\bigg[\phi+\gamma-\frac\pi{2}+\frac{{A}\cos\zeta}{\omega_0}
(\cos\phi+\sin\delta)+2m\pi \bigg]\bigg/\omega_0,
\end{equation}
where $m$ is an integer. Thus when the oscillating control field
is tuned at the resonant frequency, the total time to implement a
single-qubit operation is also quantized in this case of a 
non-perpendicular control field.

\subsection{Constant control field}
We now consider the case when the control $v$ can be varied within a
given range $[v_{0}, v_{1}]$.  A well known control strategy is
Bang-Bang control, which pertains when the control field switches back
and forth between two extremal values $v_{0}$ and
$v_{1}$~\cite{Bryson:75}.  In classical control theory, it can be
shown that Bang-Bang control is indeed the time optimal control
strategy for a double integrator system. We will show that for a
quantum system of the form in Eq.~\eqref{eq:t2}, Bang-Bang control can
generate an arbitrary single-qubit operation with the minimum number
of switchings. We note that the term Bang-Bang control has recently
been adopted with a somewhat different meaning in the study of
dynamical coupling of open quantum systems~\cite{Viola:99, Viola:02,
  Wu:02, Byrd:02}, where it is referred to performing instantaneously
or as fast as physically possible, a set of unitaries, implying that
the corresponding set of Hamiltonians can be turned on for negligible
amounts of time with (ideally) arbitrarily large strength.  A modified
form of this approach for finite control field amplitudes and
durations has been presented in Ref.~\cite{Viola:03}.

We rewrite the quantum system~\eqref{eq:t2} as follows:
\begin{equation}
  \label{eq:t26}
  i\dot U = (a\sigma_z + v H_c) U,
\end{equation}
where $v\in [v_{0}, v_{1}]$. Without loss of generality, we can assume
that $v_0=0$. By taking extremal values $v=0$ and $v=v_1$, we obtain
two linearly independent Hamiltonians 
\begin{eqnarray}
H_1&=&a\sigma_z,\\
H_2&=&a\sigma_z+v_1H_c. 
\label{eq:h1h2}
\end{eqnarray}
Our task is to achieve an arbitrary
single-qubit target operation $U_T$ by switching between 
$H_1$ and $H_2$. The final operation generated by
switching between these two Hamiltonians can be described as
\begin{equation}
\label{eqfg:2}
U_T=e^{-iH_1t_n} e^{-iH_2t_{n-1}}\cdots e^{-iH_1t_3} e^{-iH_2t_2}e^{-iH_1t_1}.
\end{equation}
In the mathematical literature, this problem is also known as the
uniform finite generation problem on Lie groups, and can be traced
back to the early 1970's~\cite{Low71, Low72}. Recently, D'Alessandro
studied the optimal evaluation of generalized Euler angles on the
rotation group with this approach~\cite{DAl02b}.

To solve this problem, we first rewrite $H_2$ as:
\begin{equation*}
H_2= a\sigma_z+v_1H_c=b_1\sigma_x+b_2\sigma_y+b_3\sigma_z.
\end{equation*}
Let $k=e^{i\sigma_z\gamma}$, where $\gamma$ satisfies
\begin{equation}
\label{eq:gamma}
 -b_1\sin 2\gamma+b_2\cos 2\gamma=0.
\end{equation}
From the Campbell-Baker-Hausdorff (CBH) formula~\cite{Warner:83}, we have:
\begin{eqnarray}
\label{eq:stan}
\aligned
\tilde H_1&=k a\sigma_z k^\dag= a\sigma_z,\\
\tilde H_2&=k H_2 k^\dag=e^{i\sigma_z\gamma}
(b_1\sigma_x+b_2\sigma_y+b_3\sigma_z) 
e^{-i\sigma_z\gamma}\\
&=(b_1\cos 2\gamma +b_2\sin 2\gamma)\sigma_x+b_3\sigma_z\\
&=b
(\sin\alpha\sigma_z+\cos\alpha\sigma_x),
\endaligned
\end{eqnarray}
where\begin{equation}
b=\sqrt{(b_1\cos 2\gamma +b_2\sin 2\gamma)^2+b_3^2}\
\end{equation}
and
\begin{equation}
  \label{eq:15}
\sin\alpha=\frac{\langle\tilde H_1, \tilde H_2\rangle}{\langle
\tilde H_1, \tilde H_1\rangle^{\frac12} \langle\tilde H_2, 
\tilde  H_2\rangle^{\frac12} }=
\frac{\tr( \tilde H_1 \tilde H_2)}{\tr( \tilde H_1^2)^{\frac12}\tr(
  \tilde H_2^2)^{\frac12} }=
\frac{\tr(H_1H_2)}{\tr(H_1^2)^{\frac12}\tr(H_2^2)^{\frac12} }.
\end{equation}
Without loss of generality, we can take $\alpha\in [-\frac\pi{2},
\frac\pi{2}]$. When the target operation $U_T$ can be generated in
Bang-Bang strategy as in Eq.~\eqref{eqfg:2}, we have
\begin{eqnarray*}
kU_Tk^\dag&=&k e^{-iH_1t_n} e^{-iH_2t_{n-1}}\cdots e^{-iH_1t_3}
e^{-iH_2t_2}e^{-iH_1t_1} k^\dag\\
&=&e^{i\tilde H_1 t_n} e^{i\tilde H_2 t_{n-1}}\cdots e^{i\tilde H_1 t_3}
e^{i\tilde H_2 t_2}e^{i\tilde H_1 t_1}.
\end{eqnarray*}
In order to generate the target quantum operation $U_T$ from the
original quantum system~\eqref{eq:t26}, we therefore only need to
generate $kU_Tk^\dag$ from the two Hamiltonians $\tilde H_1$ and
$\tilde H_2$ in Eq.~\eqref{eq:stan}.  For the special case
$\alpha=\frac{\pi}2$, we have $\tilde H_2=b\sigma_x$ and $U_T$
simplifies to the familiar Euler ZXZ decomposition of
Eq.~\eqref{eq:t10}:
\begin{equation*}
  U_T=e^{-i(\phi-\pi/2)/2\sigma_z} e^{-i(\pi-\theta)/2\sigma_x} 
e^{-i\gamma/2 \sigma_z}.
\end{equation*}

\begin{figure}[tb]
\begin{center}
\begin{tabular}{ccc}
 \psfrag{x}[][]{$x$}
  \psfrag{z}[][]{$z$}
  \psfrag{y}[][]{$y$}
\includegraphics[width=0.3\hsize]{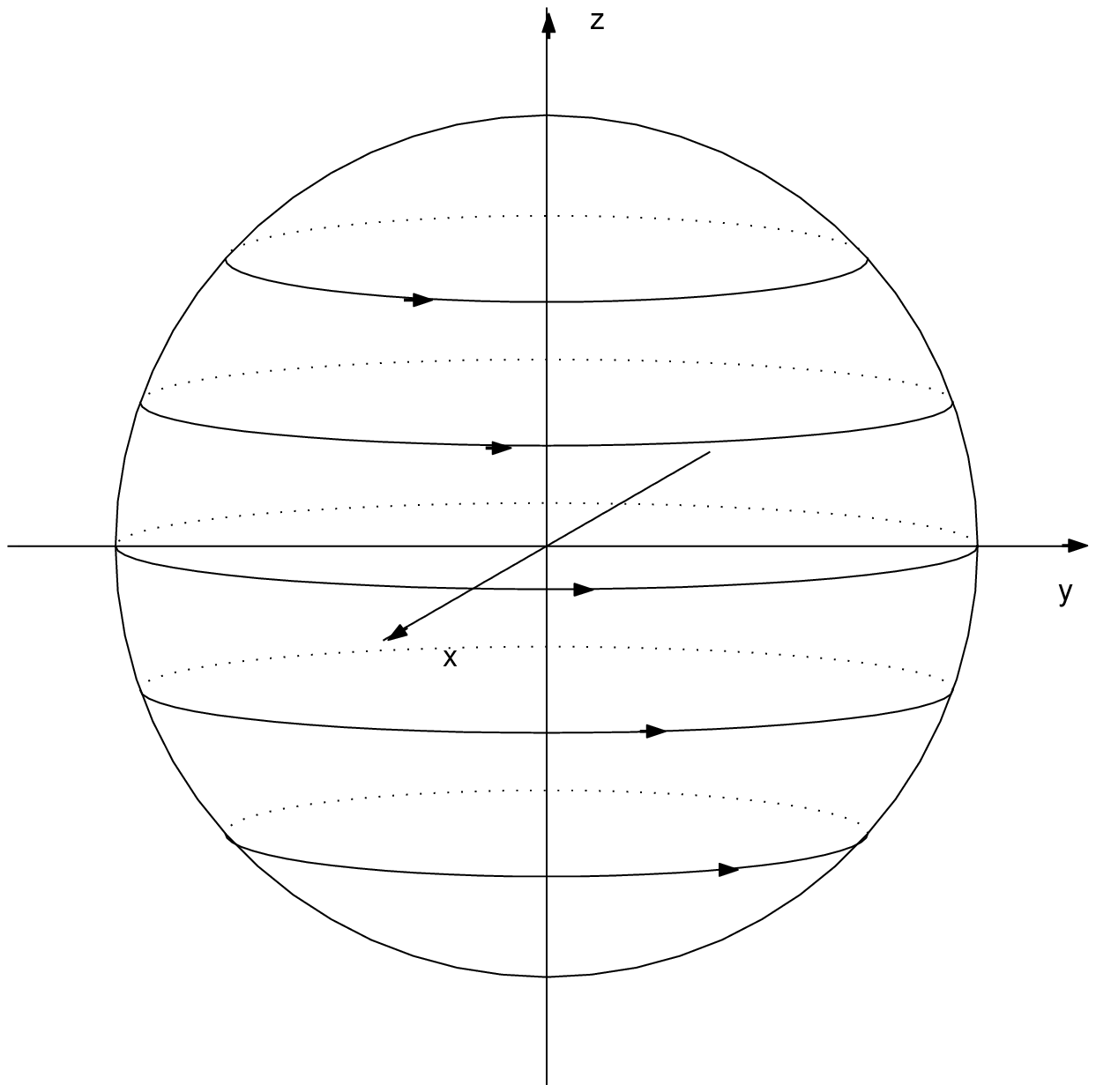} 
&\quad 
 \psfrag{x}[][]{$x$}
  \psfrag{z}[][]{$z$}
\psfrag{y}[][]{$y$}
\psfrag{a}[][]{$\alpha$}
 \includegraphics[width=0.3\hsize]{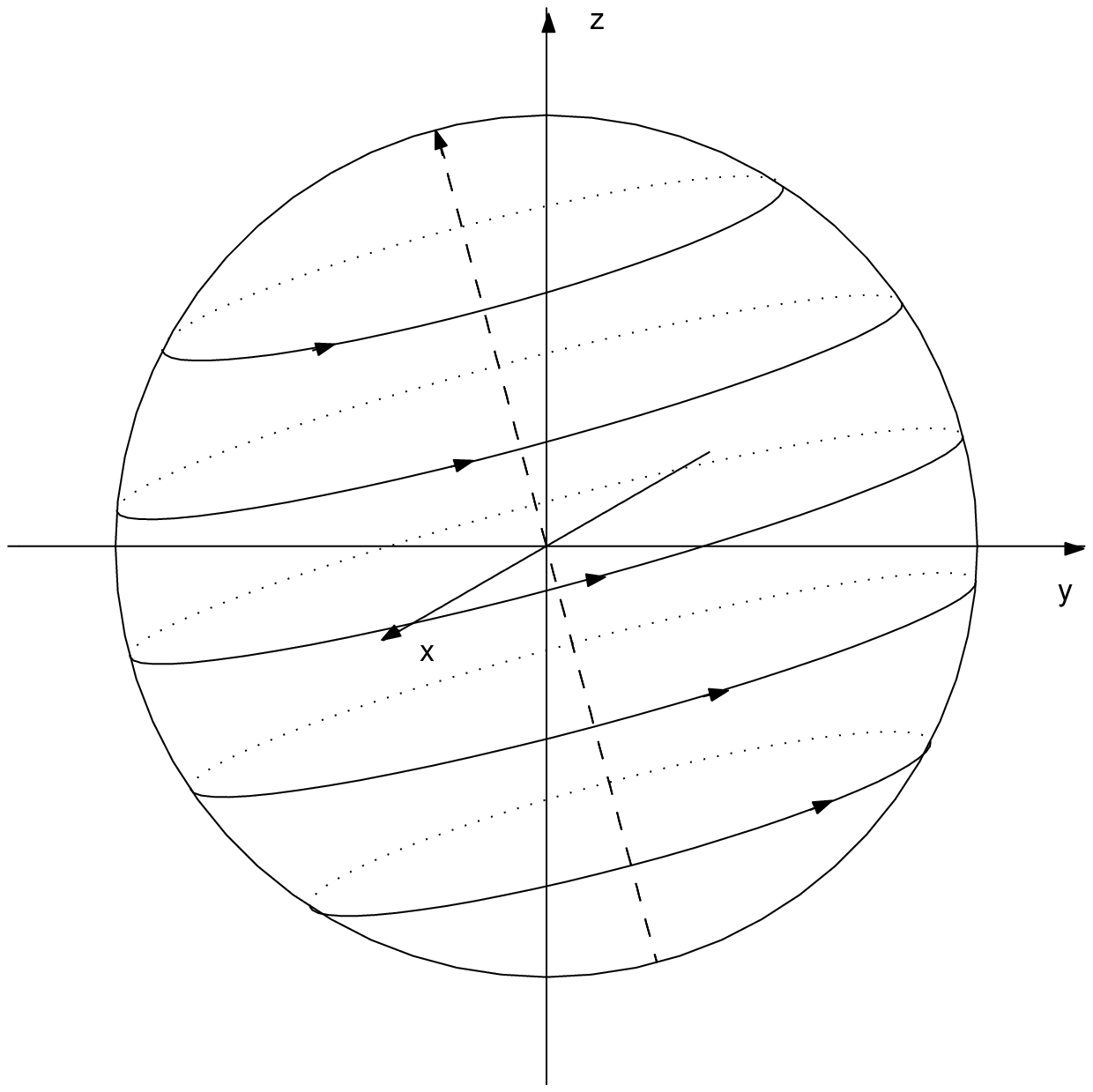}
 &\quad
 \psfrag{A1}[][]{$A_1$}
 \psfrag{A2}[][]{$A_2$}
 \psfrag{A3}[][]{$A_3$}
 \psfrag{A4}[][]{$A_4$}
 \psfrag{A5}[][]{$A_5$}
 \psfrag{A6}[][]{$A_6$}
  \psfrag{x}[][]{$x$}
   \psfrag{z}[][]{$z$}
  \psfrag{y}[][]{$\alpha$}
 \includegraphics[width=0.3\hsize]{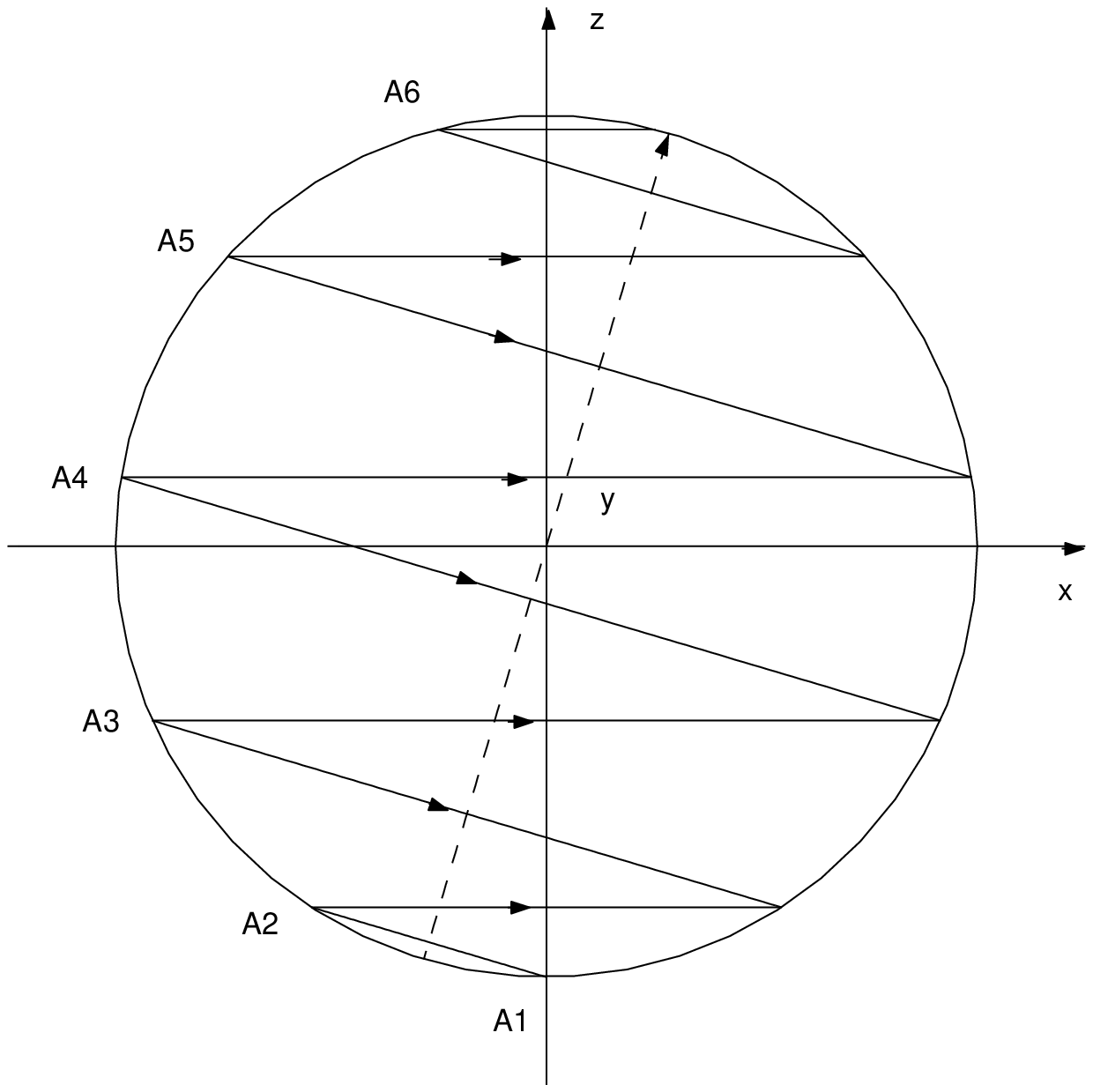} \\
 (A)&\quad(B)&\quad (C)
\end{tabular}
\end{center}
\caption{Effect of the rotations $e^{-iH_1t}$  (A) and $e^{-iH_2t}$ (B) on
  the state of a single qubit. (A) a rotation about the z-axis; (B) a
  rotation about the axis ${\hat n}=(\cos\frac\pi{6}, 0,
  \sin\frac\pi{6})$; (C) successive switching between the two
  rotations (A) and (B) applied with finite time durations allows any
  point on the Bloch sphere to be reached.}
\label{fig:steering}
\end{figure}
The Bloch sphere steering technique can now be used to find the time
sequences for switching between $H_1$ and $H_2$. We know that the
effect of $e^{-iH_1t}$ on the quantum state $|1\rangle$ is a rotation
about the $z$-axis, whereas the effect of $e^{-iH_2t}$ is a rotation
about the axis $(\sin\alpha, 0, \cos\alpha)$, as shown in
Fig.~\ref{fig:steering}(A) and (B).  Application of the time sequence
given in Eq.~\eqref{eqfg:2} to the quantum state $|1\rangle$ results
in an alternation of rotations about these two axes, illustrated in
Fig.~\ref{fig:steering}(C).  Therefore, to generate $U_T$, we need to
reach the point $U_T|1\rangle$ from the south pole $|1\rangle$ on the
Bloch sphere by switching back and forth between two rotations. It is
evident that we can now achieve any arbitrary point on the Bloch
sphere, by using these two rotations in alternation, with finite
switching times. Solution for the number of switches and time
durations is illustrated explicitly by the following example.

\begin{example}[Bang-Bang solution for Hadamard gate]
\label{ex:hadamard}
  Suppose that in the single-qubit system~\eqref{eq:t2}, we have $v\in
  [0, a\sqrt{3}]$. This gives the
  following two Hamiltonians defined by the extremal values of $v$:
\begin{eqnarray}
\label{eq:hadamard}
\aligned
H_1&=a\sigma_z,\\
H_2&=2a (\sin\frac{\pi}6\sigma_z+\cos\frac{\pi}6\sigma_x).
\endaligned
\end{eqnarray}
We aim to generate the Hadamard gate
  \begin{equation*}
    U_H=\frac1{\sqrt{2}}\left(
      \begin{matrix}
      1&1\\1&-1
      \end{matrix}
\right)
\end{equation*}
by switching between these two finite Hamiltonians at well defined times.
We first find the point corresponding to the Hadamard gate $U_H$ on the
Bloch sphere. Computing
 \begin{equation*}
  \Phi(U_H)=U_H|1\rangle =\frac1{\sqrt{2}}\left(
      \begin{matrix}
      1\\-1
      \end{matrix}
\right),
\end{equation*}
and using Eq.~\eqref{eq:hopf}, we find that $U_H$ corresponds to the
point $(\theta, \phi)=(\frac\pi{2}, \pi)$ on the Bloch sphere. The
Hamiltonians $H_1$ and $H_2$ of Eq.~\eqref{eq:hadamard} generate two
rotations on the Bloch sphere, one about the $z$-axis and the other
about the axis ${\hat n}=(\cos\frac\pi{6}, 0, \sin\frac\pi{6})$ (see
Fig.~\ref{fig:eg}).  We can implement the Hadamard gate $U_H$ by
switching between these two rotations as follows:
\begin{equation*}
  U_H=-i e^{-i H_1 t_3} e^{-i H_2 t_2} e^{-iH_1 t_1},
\end{equation*}
where $t_1=9.90/a$ and $t_2=t_3=\frac1{2a}\cos^{-1}\frac1{\sqrt{3}}$.
From Fig.~\ref{fig:eg}, it is also clear that for this Hamiltonian,
we can generate any arbitrary single-qubit operation with at most
three switchings.
\begin{figure}[tb]
\begin{center}
 \psfrag{x}[][]{$x$}
  \psfrag{z}[][]{$z$}
 \psfrag{y}[][]{$U_H$}
 \psfrag{a}[][]{$\alpha$}
\includegraphics[width=0.3\hsize]{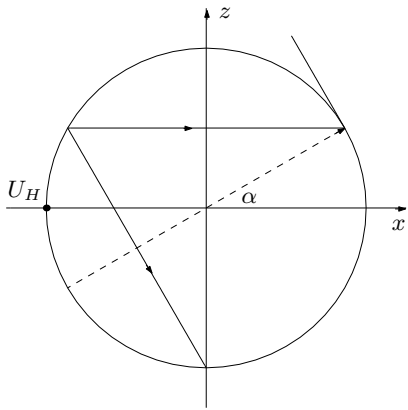} 
\end{center}
\caption{Two rotations generated by the Hamiltonians given in 
Eq.~\eqref{eq:hadamard}.}
\label{fig:eg}
\end{figure}
\end{example}

The minimum number of switchings between the two Hamiltonians $H_1$
and $H_2$ needed to generate a target quantum operation $U_T$ is
determined by the rotation axis $(\sin\alpha, 0, \cos\alpha)$ and the
$z$ coordinate of the point $U_T|1\rangle$ on the Bloch sphere. When
$\alpha=0$, we can get to any point on the sphere starting from the
south pole ($|1\rangle$) by at most two switchings. This corresponds
to the Euler ZXZ decomposition. For a general value of $\alpha \in [0,
\pi/2)$, it is not hard to derive that the coordinates for $A_n$ in
Fig.~\ref{fig:steering}(C) is $A_n=[\cos\alpha_n, -\sin\alpha_n]$,
where $\alpha_n=\pi/2+2(n-1)(\pi/2-\alpha)$. To ensure any point on
the sphere can be reached by the combination of these two rotations,
we only require the condition that $\alpha_n \ge 3\pi/2$, which in
turn yields that we can implement any arbitrary single-qubit operation
with at most $\lceil \pi/(\frac\pi{2}-\alpha)\rceil$ switchings, where
the ceiling function $\lceil x \rceil$ is defined as a function that
rounds $x$ to the nearest integer towards infinity. Similarly, when
$\alpha\in (\pi/2, \pi]$, the maximal switchings needed is $\lceil
\pi/(\alpha-\frac\pi{2})\rceil$.  To achieve the minimum number of
switchings, we therefore need to use the extremal values of the
control to minimize the angle $\alpha$.

We note that the generation of arbitrary single-qubit
operations by switching between two constant control fields has many other
applications in quantum computation and quantum information
processing. For example, in quantum simulations, if we are given two
single-qubit Hamiltonians $H_1$ and $H_2$, the usual way to generate
$U=e^{iH\Delta t}$ is to use the Trotter formula~\cite{Bennett:01,
  Dodd:02, Jane:03}
\begin{eqnarray*}
  e^{i(H_1+H_2)\Delta t}= e^{iH_1\Delta t} e^{iH_2\Delta t}+O(\Delta t^2).
\end{eqnarray*}
We can use the above Bang-Bang approach to simulate $U=e^{iH\Delta t}$
in a constructive and exact manner, without assumption of impulsive
and arbitrarily strong controls~\cite{Viola:99} or of infinitesimal
operations~\cite{DiVincenzo:95}.  Another example is in encoded
universal quantum computation with the exchange
interaction~\cite{DiV:00}.  Here the encoded single-qubit quantum
gates are generated by two Hamiltonians identical to those in
Eq.~\eqref{eq:hadamard}.  The switching sequence to implement any such
encoded single-qubit operation can readily be derived using the
strategy outlined above, which yields the result that only four
switches are necessary, corresponding to the solution found by
numerical arguments in~\cite{DiV:00}. Finally, it has been shown
recently that electron spin in a semiconductor heterostructure may be
controlled by a single-qubit Hamiltonian generated by $g$-tensor
engineering that is realized by applying a DC bias
voltage~\cite{Kato:03}. By switching between two Hamiltonians obtained
from two different $g$-factors corresponding to different bias
voltages, we can thus implement any single-qubit quantum operation.

\section{Generation of two-qubit operations: steering in the Weyl chamber}
\label{sec:two}
To generate an arbitrary quantum operation on arbitrarily many qubits,
the target quantum operation can be decomposed into a combination of
single- and two-qubit operations~\cite{Barenco:95}.  In this section we now
consider the generation of two-qubit operations. 

A general Hamiltonian for a two-qubit quantum physical system can be
written as
\begin{equation}
  \label{eq:1}
 H=g_1 \cdot  \overrightarrow{\sigma}\otimes I+I\otimes 
g_2 \cdot  \overrightarrow{\sigma}+S,
\end{equation}
where $\overrightarrow{\sigma}$ is the vector $(\sigma_x, \sigma_y,
\sigma_z)$ of Pauli matrices, and $S=\sum
J_{\alpha\beta}\sigma_\alpha^1\sigma_\beta^2$ with
$\sigma_{\alpha}^1\sigma_{\beta}^2 = \sigma_{\alpha} \otimes
\sigma_{\beta}$, $\alpha$, $\beta\in \{x, y, z\}$, and $
J_{\alpha\beta}$ the corresponding coupling strength.  The
coefficients $g_1=[g_{1x}, g_{1y}, g_{1z}]$, $g_2=[g_{2x}, g_{2y},
g_{2z}]$ may be regarded as single-qubit ``g-factors'', although they
may have very different physical origins, depending on the physical
realization of the qubits.  From~\cite{Bennett:01,Khaneja:01,
  Zhang:02}, we know that for any arbitrary set of coupling strengths
$\{J_{\alpha\beta}\}$, there exists a local operation $k$ such that
\begin{eqnarray}
  \label{eq:h1}
  kSk^\dag=J_x\sigma_x^1\sigma_x^2+J_y\sigma_y^1\sigma_y^2
+J_z\sigma_z^1\sigma_z^2.
\end{eqnarray}
Without loss of generality, we can assume for the two-qubit
steering problem that the two-body interaction term $S$ is always
given as in Eq.~\eqref{eq:h1}.  We shall study the generation of an
arbitrary quantum operation from relevant physical Hamiltonians by
tuning of the parameters $g_1$, $g_2$, and $J_x$, $J_y$, $J_z$, all of which
correspond to control parameters in this context.

\begin{figure}[t]
\begin{center}
\psfrag{c1}[][]{$c_1$}
\psfrag{c2}[][]{$c_2$}
\psfrag{c3}[][]{$c_3$}
  \psfrag{A1}[][]{$A_1$}
 \psfrag{A2}[][]{$A_2$}
 \psfrag{A3}[][]{$A_3$}
 \psfrag{O}[][]{$O$}
\psfrag{B}[][]{$B$}
 \psfrag{L}[][]{$L$}
\includegraphics[width=0.3\hsize]{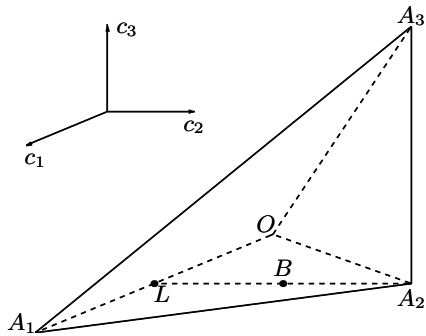} 
\end{center}
 \caption{Tetrahedral (or Weyl chamber) representation of local
   equivalence classes of nonlocal two-qubit operations from
   Ref.~\cite{Zhang:02}. Points within the tetrahedron are labeled
   $[c_1, c_2, c_3]$, where $\pi - c_2 \geq c_1 \geq c_2 \geq c_3 \geq
   0$.  Points $O([0, 0, 0])$ and $A_1([\pi, 0, 0])$ correspond to
   local operations, $L([\frac\pi{2}, 0, 0])$ to the CNOT gate, and
   $A_3([\frac\pi{2}, \frac\pi{2}, \frac\pi{2}])$ to the SWAP gate. }
\label{fig:tetra}
\end{figure}

In Ref.~\cite{Zhang:02} we have shown that this problem can be reduced
to a steering problem in a tetrahedron.  Central to this analysis is
the notion of local equivalence. Two quantum operations $U$, $U_1\in
\SU(4)$ are called {locally equivalent} if they differ only by local
operations: $U= k_1 U_1 k_2$, where $k_1$, $k_2\in \SU(2)\otimes
\SU(2)$ are both local operations. Clearly this defines an equivalence
relation on the Lie group $\SU(4)$ of all the two-qubit operations. It
was shown in~\cite{Zhang:02} that the local equivalence classes of
two-qubit operations are in one-to-one correspondence with the points
in the tetrahedron $OA_1A_2{A}_3$ as shown in Fig.~\ref{fig:tetra},
except on its base (where there is a one-to-two correspondence). This
tetrahedron is a Weyl chamber that contains all possible local
equivalence classes. To locate the point in the tetrahedron that
corresponds to a given two-qubit operation, we make use of a set of
local invariants.  Ref.~\cite{Makhlin:00} provides a simple procedure
to compute three real quantities $g_1$, $g_2$ and $g_3$ for a given
two-qubit operation and showed that two quantum operations are locally
equivalent if and only if they have identical values of these three
invariants. These local invariants can be expressed in terms of the
tetrahedron coordinates $[c_1, c_2, c_3]$ as follows~\cite{Zhang:02}:
 \begin{eqnarray}
 \label{eq:cg}
  \aligned
 g_1&=4\cos c_1\cos c_2 \cos c_3,\\
 g_2&=4\sin c_1 \sin c_2 \sin c_3,\\
 g_3&=\cos 2 c_1+\cos 2 c_2+\cos 2 c_3\\ 
 \endaligned.
 \end{eqnarray}
By solving Eqs.~\eqref{eq:cg}, we can obtain the point 
$[c_1, c_2, c_3]$ in the tetrahedron that represents a given
two-qubit operation.
 
 As time evolves, the quantum system controlled by $H$ 
generates a continuous flow in
 the space of all the quantum operations. At any time instant $t$, the
 quantum operation $U(t)$ on this flow can be mapped to a point in the
 tetrahedron. This defines a continuous
 trajectory within the Weyl chamber.  We then employ a two step
 procedure to generate a target two-qubit operation from a given
 Hamiltonian. First, we steer the Weyl chamber trajectory generated by
 the given Hamiltonian $H$ to reach the point representing the local
 equivalence class of the desired target operation. Second, we
 perform the necessary local operations that transform the quantum
 operation at this Weyl chamber point to the true target operation.
 The single-qubit operations are generated as described in 
Section~\ref{sec:single}. Therefore we only deal with the first step here, 
namely steering of
 the system Hamiltonian $H$ to generate a trajectory that hits the
 target in the tetrahedron $OA_1A_2{A}_3$.

For a general Hamiltonian containing both local and nonlocal terms, it
is usually difficult to implement an arbitrary two-qubit operation
directly. This is reflected in the complexity of a general trajectory in the 
Weyl chamber.  One approach is to introduce an intermediate step to
construct a universal gate set of elementary gates that can be used to
construct any arbitrary two-qubit operation~\cite{Barenco:95}.  An
alternative perspective is to seek to minimize the time for nonlocal
operations~\cite{Khaneja:01,Hammerer:02}.  We follow the first
approach of steering to achieve a universal gate here. The solutions will
then not necessarily be time optimal.  Optimization for time can be performed
subsequently in comparison of different gates~\cite{Zhang:04}. 

One well-known example
of such a universal gate set is the CNOT gate together with
single-qubit operations~\cite{Barenco:95}.  In~\cite{Zhang:03b}, we
showed that both CNOT and Double-CNOT (or its locally equivalent
variant, iSWAP~\cite{Ech:01,Schuch:03}) require at most three
applications to realize any two-qubit operation. More recently, we
discovered a new quantum operation B that is universal by at most two
applications together with at most six single-qubit
operations~\cite{Zhang:04}.  Whenever the direct generation of any
quantum operation is difficult, we can take advantage of these results for 
efficient
two-qubit quantum circuits and first seek to
implement the quantum gate B or CNOT, then use the corresponding
analytic circuits of ~\cite{Zhang:03b,Zhang:04} in order to construct
an arbitrary two-qubit operation.
 
To illustrate the basic idea of the steering approach, consider first
a simple case when $g_1=g_2=0$, i.e., the Hamiltonian contains only
the nonlocal term $S$.  The Weyl chamber trajectory generated by this
Hamiltonian is the straight line $U(t)=[J_x, J_y, J_z]\,
t$~\cite{Zhang:02}. The evolution direction of this line can be
changed by conjugating with an appropriate local operation from the
Weyl group.  Since the tetrahedron is a three dimensional geometric
object, it is evident that by changing the direction at most twice, we
can steer the Hamiltonian to anywhere in the tetrahedron. In other
words, by turning on the Hamiltonian at most three times, together
with application of at most four local gates, we can implement any
arbitrary two-qubit operation~\cite{Zhang:02}.  Note that when the
local terms commute with the nonlocal term, we can drop the local
terms and the trajectory is unchanged.  For example, with
$H=g_{1x}\sigma_x^1+g_{2x}\sigma_x^2+J_x\sigma_x^1\sigma_x^2$,
corresponding to the Hamiltonian derived in the superconducting qubit
proposal of ~\cite{You:02}, the Weyl chamber trajectory is exactly the
same as that generated by $J_x\sigma_x^1\sigma_x^2$, namely along the
straight line $OA_1$.  In this case, the straight line trajectory
leads directly from the origin (local unitary) to the CNOT gate
$[\frac{\pi}{2}, 0, 0]$.  This result was derived by more complex
means in Ref.~\cite{You:02}.

\subsection{Isotropic qubit coupling}
Isotropic coupling of qubits, characterized by coupling amplitudes
$J_x=J_y=J_z=J$ is a very common form of two-body interaction that
appears in many physical qubit proposals involving electronic spins,
e.g., spin-coupled coupled quantum dots~\cite{Loss:98, Burkard:99} and
donor spins in semiconductors~\cite{Kane:98, Vrijen:00, Levy:01,
  Friesen:03, Skinner:03}. We present here two different strategies to
steer the Weyl chamber trajectories generated by this Hamiltonian,
both of which are based on the ability to tune the relative values of
the single-qubit parameters $g_1$ and $g_2$.  The technological
ability to make this tuning has been demonstrated for electron spins
in semiconductors in Ref.~\cite{Kato:03}.  For greater simplicity, in
the remainder of this section we use a slightly modified notation for
the two-qubit interaction, namely the form $JS$ where
$S=\sigma_x^1\sigma_x^2+\sigma_y^1\sigma_y^2 +\sigma_z^1\sigma_z^2$
and $J$ is the isotropic coupling amplitude.

The first strategy is based on tuning $g_1$ and $g_2$ to be equal,
i.e., setting $g_1=g_2$.  Then the individual components of the total
single-qubit spins commute with the interaction term $S$,
\begin{eqnarray*}
  \label{eq:8}
[\sigma_x^1+\sigma_x^2,S]=0, \quad
[\sigma_y^1+\sigma_y^2, S]=0,\quad
\ [\sigma_z^1+\sigma_z^2, S]=0.
\end{eqnarray*}
This implies that single-qubit and two-qubit terms commute,
\begin{eqnarray*}
  \label{eq:10}
& & [g_1\cdot\overrightarrow{\sigma}\otimes I+I\otimes 
g_1 \cdot  \overrightarrow{\sigma}, S]\\
&=&[g_{1x}(\sigma_x^1+\sigma_x^2)+
g_{1y}(\sigma_y^1+\sigma_y^2)+g_{1z}(\sigma_z^1+\sigma_z^2), S]=0,
\end{eqnarray*}
allowing the time evolution to be factorized exactly:
\begin{eqnarray*}
  \label{eq:11}
  e^{iHt}=e^{i g_1\cdot\overrightarrow{\sigma}t}\otimes 
e^{i g_2\cdot\overrightarrow{\sigma}t}\cdot e^{iJSt}.
\end{eqnarray*} 
Consequently, the flow generated by the Hamiltonian $H$ in the
tetrahedron is the same as that generated by $S$, i.e., we have
achieved a reduction to the purely nonlocal interaction. The resulting
Weyl chamber trajectory lies along the line $OA_3A_1$, and we can
achieve the CNOT gate by changing the evolution direction just once
(cf. Example 4 in Ref.~\cite{Zhang:02}).  We emphasize that this
reduction results from the tuning to equivalence of the two single
qubit terms.  Furthermore, it is valid independent of the nature of
the single-qubit terms, i.e., whether they involve $\sigma_x,
\sigma_y$ and/or $\sigma_z$. Thus, although the result resembles the
known result that the isotropic exchange interaction generates a
$\sqrt{\text{SWAP}}$ gate from which a CNOT gate can be obtained by
conjugating with single-qubit gates \cite{Burkard:99b}, it is valid
here for a more general Hamiltonian that contains both nonlocal and
local terms.  The extended range of validity is a consequence of the
decomposition of the two-qubit operation into nonlocal equivalence
classes and local operations.

\begin{figure}[t]
\begin{center}
 \psfrag{A}[][]{$O$}
 \psfrag{B}[][]{$\frac\pi{4}$}
 \psfrag{C}[][]{$\frac\pi{2}$}
 \psfrag{D}[][]{$\frac{3\pi}4$}
 \psfrag{E}[][]{$\pi$}
 \psfrag{G}[][]{$\frac{\pi}4$}
 \psfrag{H}[][]{$\frac{\pi}2$}
 \psfrag{c1}[][]{$c_1$}
\psfrag{A3}[][]{$A_3$}
\psfrag{A1}[][]{$A_1$}
\psfrag{c2}[][]{$c_2$}
\psfrag{b}[][]{$B$}
\includegraphics[width=50mm]{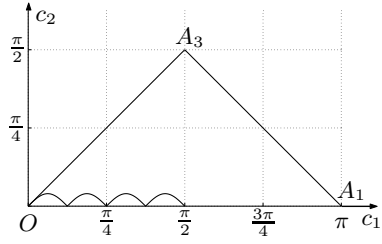}
\end{center}
\caption{Weyl chamber trajectory obtained upon steering the isotropic 
  Hamiltonian to reach CNOT by tuning the 
  local terms $g_1$ and $g_2$ to be linearly dependent.  The Weyl
  chamber trajectory is located in the $OA_3A_1$ plane of the
  tetrahedron, and goes from the origin to the point $[\pi/2, 0, 0]$
  representing the local equivalence class of CNOT, with a total of
  four oscillations.}
\label{fig:5}
\end{figure}

The second strategy is based on tuning the single-qubit parameters
$g_1$ and $g_2$ such that they become linearly dependent, i.e.,
$g_1=\lambda g_2$, where $\lambda$ describes the parameter ratio.  This 
is clearly a less stringent tuning requirement than the first strategy. 
After some mathematical analysis, the trajectory generated by the Hamiltonian 
in the tetrahedron is found to be:
\begin{eqnarray*}
  \label{eq:p1}
\aligned
  c_1&= 2 J t,\\
c_2&=c_3=|\sin^{-1}(\frac{2J}\omega \sin \omega t)|,
\endaligned
\end{eqnarray*}
where $\omega={\sqrt{(\lambda-1)^2||g_2||^2+4J^2}}$. Since $c_2=c_3$, the 
entire trajectory lies in the $OA_3A_1$ plane of the
tetrahedron $OA_1A_2A_3$, as shown in Fig.~\ref{fig:5}.  
CNOT is located on the 
intersection of this plane with the base of the tetrahedron, at $[\pi/2, 0, 0]$,
and hence
constitutes a natural target for these trajectories. 
It is readily verified that in order to steer the trajectory to CNOT we only 
need to satisfy the conditions
\begin{eqnarray}
  \label{eq:p2}
  2Jt&=&\frac\pi{2},\\
 \label{eq:p3}
\omega t&=& m\pi,
\end{eqnarray}
where $m$ is a positive integer. From these two equations, we find
that the time required to reach CNOT with this trajectory
is $t=\frac\pi{4J}$, and the parameter ratio $\lambda$ 
has to satisfy the condition
\begin{eqnarray}
  \label{eq:p5}
(\lambda-1)^2||g_2||^2=(16m^2-4)J^2.
\end{eqnarray}
From Eq.~\eqref{eq:p2}, we find that the time to achieve CNOT depends
only on the coupling strength $J$, while Eq.~\eqref{eq:p3} shows that
the integer $m$ determines the number of oscillations in the
trajectory. It is interesting to note that this time is equal to the
time needed to achieve CNOT from a purely isotropic Hamiltonian $JS$,
or from the first strategy discussed above, both of which have a
straight line Weyl chamber trajectory~\cite{Zhang:02}. 

As an example, consider a system with $J=0.1$ and choose $m=4$. Then
Eq.~\eqref{eq:p5} becomes 
\begin{equation*}
  (\lambda-1)^2||g_2||^2=2.52,
\end{equation*}
which needs to be solved for $\lambda$, given a specific set of $g$
parameters.  Choosing $g_2=[4, 4, 4]$, we obtain $\lambda=0.7709$,
$\omega=1.6$, and find a time $t=2.5\pi$ to reach CNOT.  The resulting
Weyl chamber trajectory is shown in Fig.~\ref{fig:5}. 

\subsection{Anisotropic qubit coupling}
Hamiltonians of the general form of Eq.~(\ref{eq:1}) with anisotropic
coupling coefficients $J_{\alpha \beta}$ are encountered in many
proposed solid state implementations of quantum computation.  These
include quantum dots~\cite{Imamoglu:99}, electrons coupled by long
range quantum Hall effects~\cite{Privman:98}, electrons on
helium~\cite{Platzmann:99}, and atoms in cavities~\cite{Zheng:00}.

To illustrate the Weyl chamber steering approach for anisotropic qubit
couplings, we consider here a Hamiltonian containing an Ising
interaction, i.e., a single diagonal component of $J_{\alpha \beta}$
only:
\begin{equation*}
  \label{eq:yy71}
 H_{yy}=g_{1x}\sigma_x^1+g_{1z}\sigma_z^1+g_{2x}\sigma_x^2
  +g_{2z}\sigma_z^2+J\sigma_y^1\sigma_y^2.
\end{equation*}
This Ising-type coupling between qubits is commonly seen in proposals
for superconducting Josephson junction qubits~\cite{Niskanen:03,
  Makhlin:99} and also arises as one limit of dipole-dipole or
$J$-coupling systems~\cite{Lieven:04, Twamley:03}.  It is
straightforward to show that the Hamiltonian $H_{yy}$ is locally
equivalent to the corresponding Hamiltonians $H_{xx}$ and $H_{zz}$:
\begin{eqnarray*}
\aligned
  H_{xx}&=g_{1y}\sigma_y^1+g_{1z}\sigma_z^1+g_{2y}\sigma_y^2
  +g_{2z}\sigma_z^2+J\sigma_x^1\sigma_x^2\\
 H_{zz}&=g_{1x}\sigma_x^1+g_{1y}\sigma_y^1+g_{2x}\sigma_x^2
  +g_{2y}\sigma_y^2+J\sigma_z^1\sigma_z^2. 
\endaligned
\end{eqnarray*}
Consequently the Weyl chamber trajectories generated by $H_{xx}$ and
$H_{zz}$ are the same as those generated by $H_{yy}$.  Using Makhlin's
procedure~\cite{Makhlin:00} to evaluate the local invariants of
$U(t)=e^{iH_{yy} t}$, we find
\begin{align}
\label{eq:yy83}
g_1&=\frac{(f_1^2+J^2)J^2x^2 +(f_2^2+J^2)J^2y^2+f_1^2f_2^2-J^4}
{(f_1^2+J^2) (f_2^2+J^2)},\nonumber\\
g_2&=0,\\
g_3&=\bigg(3f_1^2f_2^2-J^2(f_1^2+f_2^2)+J^4(8x^2
  y^2+3)+4J^2y^2\nonumber\\
&\cdot(f_2^2-J^2)+4J^2 x^2(f_1^2-J^2)\bigg)\bigg/
{(f_1^2+J^2) (f_2^2+J^2)},\nonumber
\end{align}
where $x=\cos\sqrt{f_2^2+J^2}\ t$, $y=\cos\sqrt{f_1^2+J^2}\ t$,
$f_1={\sqrt{g_{1x}^2+g_{1z}^2}}+{\sqrt{g_{2x}^2+g_{2z}^2}}$, and
$f_2={\sqrt{g_{1x}^2+g_{1z}^2}}-{\sqrt{g_{2x}^2+g_{2z}^2}}$.
Without loss of generality, we can assume that $J=1$ in the subsequent
analysis. 

\begin{figure}[t]
\begin{center}
 \psfrag{A}[][]{$O$}
\psfrag{B}[][]{$\frac\pi{4}$}
\psfrag{C}[][]{$\frac\pi{2}$}
\psfrag{D}[][]{$\frac{3\pi}4$}
\psfrag{E}[][]{$\pi$}
 \psfrag{G}[][]{$\frac{\pi}4$}
 \psfrag{H}[][]{$\frac{\pi}2$}
\psfrag{c1}[][]{$c_1$}
\psfrag{c2}[][]{$c_2$}
\psfrag{b}[][]{$B$}
\psfrag{A1}[][]{$A_1$}
\psfrag{A2}[][]{$A_2$}
\includegraphics[width=50mm]{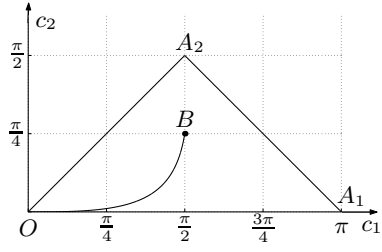}
\end{center}
\caption{Minimum time Weyl chamber trajectory that reaches the B gate from 
the Ising-type YY interaction in one switching.  See Fig.~\ref{fig:tetra} for 
definitions of the coordinates and special points.}
\label{fig:6}
\end{figure}

It turns out that both CNOT and B gates can be generated by turning on
this anisotropic Hamiltonian only once.  The B gate is more efficient
than the CNOT gate in representing arbitrary two-qubit
operations~\cite{Zhang:04} and may therefore be more useful for
quantum simulations, while the CNOT gate is required for many common
error correction protocols.  We first consider realization of the B
gate.  The local invariants are $g_1=g_2=g_3=0$~\cite{Zhang:04}.  To
find the Weyl chamber trajectory that reaches the point $B=[\pi/2,
\pi/4, 0]$ in time $t$ we need therefore only to set $g_1=g_3=0$ in
Eq.~\eqref{eq:yy83}.  This yields the following equations for $t$ that
must be satisfied simultaneously:
 \begin{eqnarray*}
   \label{eq:yy91}
   \cos 2\sqrt{f_1^2+1}\ t &=& -f_1^2 \pm \frac{\sqrt{2}}2 (f_1^2+1),\\
  \label{eq:yy92}
 \cos 2\sqrt{f_2^2+1}\ t &=&-f_2^2 \mp \frac{\sqrt{2}}2 (f_2^2+1).
  \end{eqnarray*}
  There are infinitely many solutions to these two equations,
  depending on the combination of single-qubit parameters
  $g_{i\alpha}$. Numerical analysis reveals that the time optimal solution
  for reaching the B gate is achieved by setting $f_1=1.6753$, $f_2=0$,
  with terminal time $t=\frac{3\pi}8$.  The corresponding Weyl chamber
trajectory is shown in
  Fig.~\ref{fig:6}. Note that this time optimal trajectory remains at
  all times in the basal plane of the tetrahedron $OA_1A_2A_3$. A
  similar analysis was made in~\cite{Zhang:02} for a particular instance having
  $g_{1x}=g_{2x}=-\alpha/2$, $g_{1z}=g_{2z}=0$, and $J=\alpha^2$.

 We next consider realization of the CNOT gate. The local invariants for
 the CNOT gate are $g_1=g_2=0$ and $g_3=1$~\cite{Makhlin:00}. Solving
 Eq.~\eqref{eq:yy83} we obtain the following two equations:
\begin{eqnarray*}
  \label{eq:19}
 \cos 2\sqrt{f_1^2+1}\ t &=&-f_1^2,\\
 \label{eq:20}
\cos 2\sqrt{f_2^2+1}\ t &=&-f_2^2.
\end{eqnarray*}
By numerical analysis, we find that the minimum time to achieve CNOT is
$t=3.2551$, with $f_1=0.9516$ and $f_2=0.9492$.  Comparing this with the
time optimal solution for reaching the B gate described above, it is 
clear that for such Ising-type coupling, the B gate can be generated in 
shorter time than the CNOT gate.

\subsection{Weak qubit coupling}

In many coupled qubit systems, the two-qubit interaction term is much
weaker than the local term, that is, $J$ is small in comparison to
$g_1$ and $g_2$.  For example, this is the case in circuits of flux
qubits coupled by a mutual inductance that is controlled by the
circulating current in a DC Superconducting QUantum Interference
Device (SQUID)~\cite{Britton:04}. It is also the case in most NMR
systems of exchange-coupled nuclear spins~\cite{Lieven:04}. Estimates
of parameters for exchange-coupled electron spins in quantum dots also
put these systems in this weak coupling regime at magnetic field
strengths relevant to experiments ~\cite{Burkard:99, Salis:01,
  Sousa:01, Sousa:03}. In such situations we have found that the Weyl
chamber trajectory can be approximated to good accuracy by some simple
curves within the tetrahedron. This provides a useful strategy in many
physical systems, where instead of attempting to control the exact
time evolution, we seek rather to steer only the approximating curves
in the Weyl chamber and may wish to incorporate additional factors
into numerical optimization protocols.

We first consider a simple case, namely, tuning $g_1$ and $g_2$ such
that their norms are equal, i.e., $||g_1||=||g_2||$.  In this case,
when in addition $J \ll ||g_i||$, the trajectory generated in the
tetrahedron can be well approximated by a straight line. Specifically,
when the Hamiltonian is given in the Ising form
\begin{eqnarray*}
 \label{eq:yy93}
 H=g_1 \cdot  \overrightarrow{\sigma}\otimes I+I\otimes 
g_2 \cdot  \overrightarrow{\sigma}+J_z\sigma_z^1\sigma_z^2
\end{eqnarray*}
with $||g_1||=||g_2||$ and $J_z$ small, the resulting Weyl
chamber trajectory can be approximated by the following straight lines:
\begin{eqnarray*}
[c_1, c_2, c_3]=\left\{
\begin{array}{cl}
\dfrac{J_zt}{||g_1||^2}
[2g_{1z}g_{2z},\rho, \rho],&\text{if  } 2g_{1z}g_{2z}\ge \rho;\\ & \\
 \dfrac{J_zt}{||g_1||^2}[\rho, \rho, 2g_{1z}g_{2z}], &\text{if
} 2g_{1z}g_{2z}< \rho,
\end{array}\right.
\end{eqnarray*}
where $\rho=\sqrt{g_{1x}^2+g_{1y}^2} \sqrt{g_{2x}^2+g_{2y}^2}$. Recall
that in the case when the Hamiltonian is purely nonlocal, i.e., it
contains no single-qubit terms, the Weyl chamber trajectory is a
straight line and we can generate any arbitrary two-qubit operation by
applying local unitaries to change the evolution direction of the
trajectory. We can use the same technique here to steer the Weyl
chamber trajectory to arrive at any point in the tetrahedron with at
most two switchings.

\begin{figure}[t]
\begin{center}
\psfrag{A}[][]{$O$}
 \psfrag{B}[][]{$\frac\pi{4}$}
 \psfrag{C}[][]{$\frac\pi{2}$}
\psfrag{D}[][]{$\frac{3\pi}4$}
 \psfrag{E}[][]{$\pi$}
 \psfrag{GG}[][]{$\frac{\pi}{800}$}
 \psfrag{HH}[][]{$\frac{\pi}{400}$}
 \psfrag{c1}[][]{$c_1$}
\psfrag{c2}[][]{$c_2$}
\includegraphics[width=50mm]{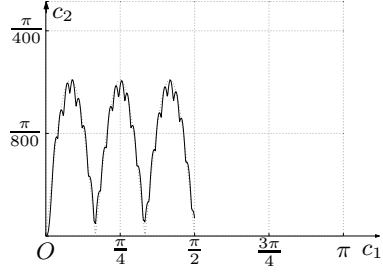}
\end{center}
\caption{ Weyl chamber trajectory designed to reach CNOT and
generated by the weak two-qubit interaction 
given in  Eq.~\eqref{eq:yy98}. The trajectory is at 
all times in or very close to the $OA_3A_1$ plane of the 
tetrahedron $OA_1A_2A_3$ of  Fig.~\ref{fig:tetra}.}
\label{fig:7}
\end{figure}

For the general case when $||g_1||\neq ||g_2||$ and
$S=J_x\sigma_x^1\sigma_x^2+J_y\sigma_y^1\sigma_y^2
+J_z\sigma_z^1\sigma_z^2$, we find that the trajectory can be well
approximated by the following sinusoidal curve in the tetrahedron:
\begin{eqnarray}
  \label{eq:yy95}
\aligned
  c_1&=\frac{J_xg_{1x}g_{2x}+J_yg_{1y}g_{2y}+J_zg_{1z}g_{2z}}
{||g_1||\cdot ||g_2||} 2t,\\
c_2&=c_3=p(g_1, g_2) |\sin (||g_1||-||g_2||)t |,
\endaligned
\end{eqnarray}
where $p$ is a nonzero function of $g_1$ and $g_2$. From
Eq.~\eqref{eq:yy95}, the approximate curve stays in the plane
$OA_3A_1$ and moves away from the origin in the direction of CNOT.
Therefore, for this Hamiltonian, the CNOT gate is a natural choice of
target operation. To reach CNOT, we require that Eq.~\eqref{eq:yy95}
be satisfied with values $c_1=\frac\pi{2}$ and $c_2=c_3=0$. It is
clear that the form of the function $p$ is irrelevant. Consequently,
we need only to satisfy the following two equations:
\begin{eqnarray}
  \label{eq:yy96}
  \aligned
\frac{J_xg_{1x}g_{2x}+J_yg_{1y}g_{2y}+J_zg_{1z}g_{2z}}
{||g_1||\cdot ||g_2||} 2t&=\frac\pi{2},\\
(||g_1||-||g_2||)t&=m\pi,
\endaligned
\end{eqnarray}
where $m$ is an integer. To illustrate this approach, consider the
following example of weak coupling:
\begin{eqnarray}
  \label{eq:yy98}
 H=g_1 \cdot  \overrightarrow{\sigma}\otimes I+I\otimes 
g_2 \cdot  \overrightarrow{\sigma}+0.2\sigma_z^1\sigma_z^2.
\end{eqnarray}
Here the local terms $g_1$ and $g_2$ are one order larger than the
interaction strength $J_z=0.2$.  From Eq.~\eqref{eq:yy96}, we
find that the local terms $g_1$ and $g_2$ must satisfy
\begin{eqnarray}
 \label{eq:yy97}
  \frac{0.2g_{1z}g_{2z}}{||g_1||\cdot ||g_2||\cdot(||g_1||-||g_2||)}
=\frac1{4m}.
\end{eqnarray}
Setting, e.g., $m=3$, we derive one solution for Eq.~\eqref{eq:yy97}
as $g_1=[2.5, 0, 10.0182]$ and $g_2=[2, 0, 7.8177]$. After a time
duration $t=4.1778$, this Hamiltonian can achieve a gate located at
$[0.5000, 0.0002, 0.0002]\pi$, which is very close to the CNOT gate.
The Weyl chamber trajectory is shown in Fig.~\ref{fig:7}, where the
solid line is the real trajectory generated by the Hamiltonian, and
the dotted line is the approximate sinusoidal curve.  A realistic
application of this example was recently made in
Ref.~\cite{Britton:04} to inductively coupled flux qubits for which
$||g_1||\neq ||g_2||$ and $J_x=J_y=0$.

\section{Conclusions}
We have developed a systematic approach to generate arbitrary
two-qubit quantum operations by steering the unitary evolution
corresponding to local and nonlocal interactions.  The generation of
single-qubit quantum gates was shown to map onto a steering problem on
the Bloch sphere.  Local two-qubit gates corresponding to coupled
single-qubit gates can then be generated by combined steering on two
coupled Bloch spheres.  The nonlocal components of two-qubit gates
were shown to be analyzable as a steering problem in a Weyl chamber of
local equivalence classes.  We applied this steering approach here to
the generation of two-qubit quantum gates in several physical examples
relevant to solid state implementations of quantum information
processing.  The methods described here are expected to provide useful
guidance to experimental design of circuits and pulse sequences for
realization of quantum logic gates.

\begin{acknowledgments}
  We thank the NSF for financial support under ITR Grant No.
  EIA-0205641, and DARPA and ONR under Grant No. FDN0014-01-1-0826 of
  the DARPA SPINs program.
\end{acknowledgments}

\bibliographystyle{apsrev} 

\end{document}